\documentclass[%
 reprint,
 amsmath,amssymb,
 aps,prd
floatfix,
showkeys
]{revtex4-1}
\usepackage{xcolor}
\usepackage{ulem}
\usepackage{graphicx}
\usepackage{dcolumn}
\usepackage{bm}
\usepackage{hyperref}

\usepackage{float}

\begin{document}


\title{Chiral symmetry restoration in a rotating medium}

\author{I. I. Gaspar}
\affiliation{Departamento de F\'isica, Universidad Aut\'onoma Metropolitana-Iztapalapa, Av. San Rafael Atlixco 186, C.P, CdMx 09340, Mexico.}

\author{L. A. Hernández}%
\affiliation{Departamento de F\'isica, Universidad Aut\'onoma Metropolitana-Iztapalapa, Av. San Rafael Atlixco 186, C.P, CdMx 09340, Mexico.}

\author{R. Zamora}
\affiliation{Instituto de Ciencias B\'asicas, Universidad Diego Portales, Casilla 298-V, Santiago, Chile.}
\affiliation{Centro de Investigaci\'on y Desarrollo en Ciencias Aeroespaciales (CIDCA), Academia Politécnica Aeronáutica, Fuerza A\'erea de Chile, Casilla 8020744, Santiago, Chile.}


\begin{abstract}
We study the nature of the chiral symmetry restoration within the Yukawa model with spontaneous symmetry breaking. We work with scalar and fermion fields which are subject to the effects of a rotating system. In this work, we show the derivation of the scalar field propagator in a rotating medium using the Fock-Schwinger proper-time method. We compute analytically the effective potential in the high-temperature approximations, including the contribution of the ring diagrams to account for the plasma screening properties. We study the chiral transition as we vary the angular velocity $\Omega$, the boson self-coupling $\lambda$ and the fermion-boson coupling $g$. We show that the critical temperature for the restoration of chiral symmetry always starts with decreasing behaviour, until it reaches a minimum and from there when increasing $\Omega$, we observe $T_c$ increases monotonically. In all the phase transition lines in the $T-\Omega$ plane reported, we obtain that the rotating effects are able to change the order of the phase transition.
\end{abstract}

\maketitle


\section{\label{sec:level1}Introduction}

In the last decades the study of the strongly interacting matter in extreme conditions has drawn a lot of attention from the high energy physics community. One of the most relevant topics to study is the phase transition that occurs in this kind of systems and conditions which are encoded in the Quantum Chromodynamics (QCD) phase diagram~\cite{Fukushima:2010bq,Andronic:2017pug,Ratti:2022qgf}. This phase diagram typically is depicted in the temperature-baryon chemical potential plane ($T - \mu_B$ plane). However, there are some other physical variables relevant when the strong phase transition happens, among which are the isospin chemical potential $\mu_I$ and the magnetic field strength, to mention a few. The reason behind all of the thermodynamical and macroscopic variables that are used to build the QCD phase diagram is the compelling need to link the theoretical results with the experimental data from different systems, such as relativistic heavy-ion collisions~\cite{Bzdak:2019pkr,Busza:2018rrf,Luo:2017faz,Braun-Munzinger:2015hba}, dense astronomical objects~\cite{Most:2018eaw,Alford:2013aca,Kurkela:2009gj} and the early universe~\cite{Mazumdar:2018dfl}. Specifically, in neutron stars the relevant variables are $\mu_B$, $\mu_I$ and $T$. Therefore, there are  works based on effective theories, Dyson-Schwinger equations, sum rules, chiral perturbation theory and the holographic method, which have reported results at high baryon and/or isospin densities and finite temperatures~\cite{Fukushima:2008wg,McLerran:2007qj,Herbst:2010rf,Fu:2019hdw,Fischer:2018sdj,Shi:2016koj,Xin:2014ela,Lopes:2021tro,Ayala:2023cnt,Mu:2010zz,Cohen:2015soa,Toublan:2003tt,Barducci:2004tt,Xia:2013caa,Mukherjee:2006hq,Stiele:2013pma,Benic:2015pia,Costa:2008yh,Gao:2016qkh,Ayala:2021tkm,Ayala:2019skg,Ayala:2017ucc,Ayala:2014jla,Ayala:2011vs,Knaute:2017opk,Loewe:2021mdo,Dominguez:2020sdf,Loewe1,Loewe2,Loewe3,Loewe4,Loewe5,Loewe6,Loewe7,Loewe8,Loewe9,Loewe10,Loewe11}. In a similar way, Lattice QCD (LQCD) has reported important result in both at finite baryon and isospin chemical potentials~\cite{Kogut:2002zg,Bazavov:2017dus,Brandt:2017oyy,Borsanyi:2012cr,Endrodi:2011gv,DElia:2002tig,Borsanyi:2021sxv,Guenther:2020jwe,Bonati:2015bha,Bellwied:2015rza}, even when the former one is technically difficult to perform as a consequence of the severe sign problem~\cite{Nagata:2021ugx}. Other important physical variable in the QCD phase transition is the magnetic field strength which is relevant when the reaction of a relativistic heavy-ion collision is studied. It is well-known that the most intense magnetic fields in the universe are produced in this kind of collisions~\cite{Brandenburg:2021lnj,Skokov:2009qp}, catalysing the deconfinement/chiral symmetry restoration. Such phenomenon is called Inverse Magnetic Catalysis (IMC)~\cite{Bali:2011qj,Bali:2012zg,Bali:2014kia}. Therefore, LQCD and effective models have reported in recent times important results including magnetic field effects in the QCD phase transition~\cite{Bali:2011qj,Endrodi:2015oba,DElia:2021yvk,Mizher:2010zb,Andersen:2014xxa,Farias:2014eca,Ferreira:2013tba,Costa:2015bza,Farias:2016gmy,Andersen:2021lnk,Bandyopadhyay:2020zte,Ayala:2021nhx,Ayala:2015lta,Ayala:2014iba,Ayala:2014gwa,Ayala:2020rmb}. Another interesting feature of the reaction in relativistic heavy-ion collisions happens when the collisions are not central. Then, the asymmetry of the matter distribution in the transverse plane causes the colliding region to develop an orbital angular velocity $\Omega$ directed along the normal to the reaction plane~\cite{Becattini:2007sr,Becattini:2015ska}, with angular velocity of $\Omega\sim 10^{22}$ s$^{-1}$~\cite{STAR:2017ckg,Jiang:2016woz}. One way to observe the effect of the high vorticity in the reaction is to analyse the global hadron polarization. Recent measurements of the global $\Lambda$ and $\overline{\Lambda}$ polarization as function of collision energy~\cite{STAR:2017ckg,STAR:2007ccu,STAR:2018gyt} show that the $\overline{\Lambda}$ polarization rises more steeply than the $\Lambda$ polarization when the collision energy decreases, theoretical determinations have reported complementary results in agreement~\cite{Xie:2019wxz,Ivanov:2020udj,Ayala:2020soy,Ayala:2021xrn}. Nevertheless, other consequence of this rotating system could be changes observed in the pseudo-critical temperature when the system shows a phase transition. This effect has been studied recently in Ref.~\cite{Chernodub:2016kxh,Jiang:2016wvv,Zhang:2020hha,Jiang:2021izj,Braga:2022yfe,Zhao:2022uxc,Chen:2022smf} and also when magnetic fields are included~\cite{Sadooghi:2021upd,Mehr:2022tfq}.

In this work, we study the phase transition of a hot and rotating system within the Yukawa model, where we include both boson and fermion fields. The rotating effect is included in the propagators of the field and we consider a thermal bath through the imaginary time formalism. The Yukawa model implements the idea of the chiral symmetry spontaneously broken, where the order parameter is the \textit{vacuum expectation value} ($vev$). It allows to follow the behaviour of the $vev$ as a function of the temperature $T$ and angular velocity $\Omega$ once the effective potential beyond the mean field theory is computed, and thus it is employed to identify the chiral symmetry restoration. 

The work is organised as follows: In Sec.~\ref{sec2}, we present the computation of the scalar field propagator in a rotating system. We use the Fock-Schwinger proper-time formalism, following the method introduced in Refs.~\cite{Iablokov:2019rpd,Iablokov:2020upc} that requires knowledge of the explicit solution of the Klein-Gordon equation. We find the solutions to the Klein-Gordon equation for bosons rigidly rotating inside a cylinder. In order to satisfy the causality condition for a given $\Omega$, the solutions are taken as not existent for $r>R$, with $R$ the cylinder radius $R$, thus $R\Omega<1$. At the end of this section with the wave-function at hand, we write the expression for a real scalar field propagator in momentum space. In Sec.~\ref{sec3}, we describe the features of the Yukawa model and then we compute the effective potential up to ring diagrams contributions, we stabilise the vacuum of our theory by introducing counter-terms to enforce that the tree-level structure of the effective potential is preserved by vacuum loop corrections. As a consequence of including ring diagrams, we compute the boson's self-energy up to 1-loop order. We work in the scenario where the medium is certainly hot, the high temperature approximation, which means the temperature is the highest energy scale in the analysis. This approximation allows us to get an analytic expression for both effective potential and self-energy. In Sec.~\ref{sec4}, we explore the parameter space looking for the critical temperature as a function of angular velocity. Since the model has three free parameters, the couplings and the squared mass parameter, we explore different possible combinations among these parameters in order to show all the possible scenarios when the chiral symmetry is restored in presence of a rotating system. In Sec.~\ref{sec5}, we summarise and conclude. Finally, the explicit computation of the one-loop boson and fermion contributions to the effective potential as well as the boson's self-energy are reported in the appendix.

\section{\label{sec2}Scalar propagator in a rotating medium}

In order to explore the phase transition associated to the chiral symmetry restoration in a system at finite temperature and rotating. We start this work by computing the corresponding scalar field propagator. The method we are going to use is known as the Fock-Schwinger proper-time method and it tells us that the two-point Green's function $D(x,x')$ is the solution of the equation
\begin{equation}
    \hat{H}(\partial_x,x)D(x,x')=\delta^4(x-x'),
\end{equation}
where $\hat{H}(\partial_x,x)$ is the Hamiltonian operator, which typically is a polynomial in $\partial_x$. The two-point Green's function can be represented as 
\begin{equation}
    D(x,x')=-i\int_{-\infty}^0 d\tau\ U(x,x';\tau),
    \label{represented}
\end{equation}
where $\tau$ is known as a proper-time parameter and $U(x,x';\tau)$ is an evolution operator in this proper-time. This operator satisfies
\begin{eqnarray}
i\partial_\tau U(x,x';\tau)=\hat{H}(\partial_x,x)U(x,x';\tau),
\label{evolutionop}
\end{eqnarray}
together with the boundary conditions
\begin{align}
U(x,x';-\infty)&=0,\nonumber\\
U(x,x';0)&=\delta^4(x-x'),
\label{boundcond}
\end{align}
where the solution is readily found as
\begin{eqnarray}
U(x,x';\tau)=\exp[-i\tau \hat{H}(\partial_x,x)]\delta^4(x-x').
\label{explU}
\end{eqnarray}
In order to find the explicit form of the proper-time evolution operator, we can use that, when the eigenfunctions $\Phi_\lambda(x)$ of the operator $\hat{H}(\partial_x,x)$ are known, the Dirac delta-function can be expressed in terms of the closure relation obeyed by the eigenfunctions $\Phi_\lambda(x)$, namely
\begin{eqnarray}
\sum_\lambda\Phi_\lambda(x)\Phi^\dagger_\lambda(x')=\delta^4(x-x'),
\label{closure}
\end{eqnarray}
where the sum over $\lambda$ represents the sum over all the quantum numbers. Consequently, an exact expression for the proper-time evolution operator can be written as
\begin{eqnarray}
U(x,x';\tau)=\sum_\lambda\exp[-i\tau\lambda]\Phi_\lambda(x)\Phi^\dagger_\lambda(x'),
\label{evolexpl}
\end{eqnarray}
where we have used the eigenvalue equation
\begin{eqnarray}
\hat{H}(\partial_x,x)\Phi_\lambda(x)=\lambda\Phi_\lambda(x).
\label{eigeneq}
\end{eqnarray}
Using Eqs.~(\ref{represented}) and~(\ref{evolexpl}), the propagator $D(x,x')$ can be written as
\begin{eqnarray}
\!\!\!\!\!\!\!D(x,x')=-i \int_{-\infty}^0d\tau\sum_\lambda\exp[-i\tau\lambda]\Phi_\lambda(x)\Phi^\dagger_\lambda(x').
\label{propG}
\end{eqnarray}

At this point, it is easy to see that we need to know the expression of the eigenfunctions $\phi_\lambda(x)$ which can be obtained from the corresponding Klein-Gordon equation. Following the path described in Ref.~\cite{Ayala:2021osy}, we find the equation which takes into account a relativistic rotating frame, where the system can be thought of as a rigid cylinder rotating around the $\hat{z}$-axis with constant angular velocity $\Omega$. It is given by
\begin{align}
    \bigg [ &\Big(i \frac{\partial}{\partial t}-\Omega i\frac{\partial}{\partial \phi} \Big)^2+\frac{\partial^2}{\partial r^2}\nonumber \\
    &+\frac{1}{r}\frac{\partial}{\partial r}+\frac{1}{r^2}\frac{\partial^2}{\partial \phi^2}+\frac{\partial^2}{\partial z^2} \bigg ] \Phi(x)= m^2 \Phi(x),
    \label{KGequation}
\end{align}
where $m$ is the mass of the scalar field and $-i\partial/\partial \phi=\hat{L}_z$. Thus, Eq.~(\ref{KGequation}) becomes
\begin{align}
        \bigg [ &\Big(i \frac{\partial}{\partial t}+\Omega \hat{L}_z \Big)^2+\frac{\partial^2}{\partial r^2}\nonumber \\
    &+\frac{1}{r}\frac{\partial}{\partial r}+\frac{1}{r^2}\frac{\partial^2}{\partial \phi^2}+\frac{\partial^2}{\partial z^2} \bigg ] \Phi(x)= m^2 \Phi(x).
    \label{KGequation2}
\end{align}
Assuming that Eq.~(\ref{KGequation2}) allows separation of variables, we propose the following solution
\begin{equation}
    \Phi[t,r,\phi,z]=e^{-iEt+ik_z z}\varphi(r,\phi).
    \label{solution1}
\end{equation}
We notice that angular momentum conservation implies to get the quantum number $l$. Therefore, the radial Klein-Gordon equation can be written as
\begin{align}
    \bigg [&(E+l\Omega)^2+\frac{\partial^2}{\partial r^2}+\frac{1}{r}\frac{\partial}{\partial r}\nonumber \\
    &-\frac{l^2}{r^2}-k_z^2-m^22 \bigg]\varphi(r,\phi)=0
    \label{KGrphi}.
\end{align}
In order to solve the radial equation, we first define the transverse momentum $k_\perp^2=\tilde{E}^2-k_z^2-m^2$ with $\tilde{E}^2=(E+l\Omega)^2$, and we get
\begin{equation}
    \bigg [ \frac{\partial^2}{\partial r^2}+\frac{1}{r}\frac{\partial}{\partial r}-\frac{l^2}{r^2}+k_\perp^2 \bigg]U(r)=0.
    \label{radialeq}
\end{equation}
We can multiply Eq.~(\ref{radialeq})  by $r^2$ and the corresponding equation is 
\begin{equation}
    \bigg [ r^2\frac{\partial^2}{\partial r^2}+r\frac{\partial}{\partial r}+(r^2k_\perp^2-l^2) \bigg]U(r)=0,
    \label{radialeq2}
\end{equation}
where it takes, after simple re-scaling $\rho=rk_\perp$, the following form
\begin{equation}
    \bigg[ \rho^2 \frac{\partial^2}{\partial \rho^2}+\rho \frac{\partial}{\partial \rho}+(\rho^2-l^2) \bigg]U(\rho)=0.
    \label{Besseleq}
\end{equation}
Equation~(\ref{Besseleq}) is the well-known Bessel equation, which solution is given by the Bessel function of the first kind $J_l(\rho)=J_l(rk_\perp)$. Therefore, the full solution to the Klein-Gordon is 
\begin{equation}
    \Phi(x)=e^{-iEt+ik_z z+i\phi l}J_l(rk_\perp).
    \label{fullsolution}
\end{equation}

Once we have Eq.~(\ref{fullsolution}) at hand, we are now able to write the Green's two point function by substituting Eq.~(\ref{fullsolution}) and its corresponding complex conjugate function in Eq.~(\ref{propG}), such that we write $D(x,x')$ as follows
\begin{widetext}
\begin{equation}
    D(x,x')=-i\int_{-\infty}^0 d\tau  \int \frac{dEdk_zdk_\perp k_\perp}{(2\pi)^3}e^{-i\tau(\tilde{E}^2-k_\perp^2-k_z^2-m^2+i\epsilon)}
     \sum_{l=-\infty}^\infty e^{-i\tilde{E}(t-t')}e^{ik_z(z-z')}e^{il(\phi-\phi')}J_l(rk_\perp)J_l(r'k_\perp).
    \label{Dexplicit}
\end{equation}
\end{widetext}
We integrate the Eq.~(\ref{Dexplicit}) over $\tau$
\begin{align}
    D(x,x')&=\sum_{l=-\infty}^\infty \int \frac{dEdk_zdk_\perp k_\perp}{(2\pi)^3} J_l(rk_\perp)J_l(r'k_\perp) \nonumber \\
    &\times \frac{e^{-i\tilde{E}(t-t')}e^{ik_z(z-z')}e^{il(\phi-\phi')}}{\tilde{E}^2-k_\perp^2-k_z^2-m^2+i\epsilon} .
\end{align}
The next step is to perform the sum over $l$. In this case, it is better to rewrite one of the Bessel function in its integral representation
\begin{equation}
    J_l(x)=\frac{1}{2\pi}\int_{-\pi}^\pi ds e^{i(x\sin{s}-ls)},
\end{equation}
then we get
\begin{align}
    D(x,x')&= \int \frac{dEdk_zdk_\perp k_\perp}{(2\pi)^3} \frac{e^{-iE(t-t')}e^{ik_z(z-z')}}{\tilde{E}^2-k_\perp^2-k_z^2-m^2+i\epsilon} \nonumber \\
    &\times \frac{1}{2\pi}\int_{-\pi}^\pi ds e^{i k_\perp r' \sin{s}}\sum_{l=-\infty}^\infty J_l(k_\perp r)\nonumber \\
    &\times e^{-i l s+il(\phi-\phi')-il\Omega(t-t')}.
    \label{propagatorBesselint}
\end{align}
If we remember the Jacobi-Anger expansion
\begin{equation}
    \sum_{l=-\infty}^\infty J_l(x)e^{il y}=e^{ix\sin{y}},
    \label{J-Aexpansion}
\end{equation}
then we are able to write $D(x,x')$ as
\begin{align}
    D(x,x')&= \int \frac{dEdk_zdk_\perp k_\perp}{(2\pi)^3} \frac{e^{-iE(t-t')}e^{ik_z(z-z')}}{\tilde{E}^2-k_\perp^2-k_z^2-m^2+i\epsilon} \nonumber \\
    &\times \frac{1}{2\pi}\int_{-\pi}^\pi ds e^{ik_\perp r \sin{((\phi-\phi')-s-\Omega(t-t'))}} \nonumber \\
    &\times e^{ik_\perp r' \sin{s}}.
\end{align}
Since we have constant angular velocity, we observe that $\Omega=(\phi-\phi')/(t-t')$, hence we obtain
\begin{align}
    D(x,x')&= \int \frac{dEdk_zdk_\perp k_\perp}{(2\pi)^3} \frac{e^{-iE(t-t')}e^{ik_z(z-z')}}{\tilde{E}^2-k_\perp^2-k_z^2-m^2+i\epsilon} \nonumber \\
    &\times \frac{1}{2\pi}\int_{-\pi}^\pi ds e^{ik_\perp r \sin{s}} e^{-ik_\perp r' \sin{s}}.
\end{align}
We proceed to make the change of variable $r'=R-h/2$ and $r=R+h/2$ and the propagator is
\begin{align}
    D(x,x')&=\int \frac{dEdk_zdk_\perp k_\perp}{(2\pi)^3} \frac{e^{-iE(t-t')}e^{ik_z(z-z')}}{\tilde{E}^2-k_\perp^2-k_z^2-m^2+i\epsilon} \nonumber \\
    &\times \frac{1}{2\pi}\int_{-\pi}^\pi ds e^{ik_\perp (R-h/2) \sin{s}} e^{-ik_\perp (R+h/2) \sin{s}}\nonumber \\
    &=\int \frac{dEdk_zdk_\perp k_\perp}{(2\pi)^3} \frac{e^{-iE(t-t')}e^{ik_z(z-z')}}{\tilde{E}^2-k_\perp^2-k_z^2-m^2+i\epsilon} \nonumber \\
    &\times \frac{1}{2\pi}\int_{-\pi}^\pi ds e^{-ik_\perp h \sin{s}}.
    \label{Dhs}
\end{align}
We notice the integral over $s$, in Eq.~(\ref{Dhs}), is nothing more than $2\pi J_0(k_\perp r)$. Besides, if we write the propagator in terms of the relative coordinates,  
 $t-t'\rightarrow t$ and $z-z'\rightarrow z$, we get the expression in coordinate space
\begin{equation}
    D(x,x')=\int \frac{dEdk_zdk_\perp k_\perp}{(2\pi)^3} \frac{e^{-iEt+ik_z z} J_0(k_\perp r)}{\tilde{E}^2-k_\perp^2-k_z^2-m^2+i\epsilon} .
\end{equation}

Finally, we introduce the Fourier transform in order to get the expression of the propagator in momentum space
\begin{align}
    D(p)&= \int \frac{dEdk_zdk_\perp k_\perp}{(2\pi)^3} \int \ dt \ dz \ d\phi \ rdr \ e^{ip_0t} \nonumber \\
    & \times e^{-i\vec{p}_\perp \cdot \vec{x}_\perp} e^{-ip_z z} \frac{e^{-iEt+ik_z z} J_0(k_\perp r)}{\tilde{E}^2-k_\perp^2-k_z^2-m^2+i\epsilon} \nonumber \\
    &= \int \frac{dEdk_zdk_\perp k_\perp}{(2\pi)^3} \int \ dt \ dz \ d\phi \ rdr \ e^{it(p_0-E)}\nonumber\\
    &\times e^{-iz(p_z-k_z)}e^{-ip_\perp r\cos{\phi}} \frac{ J_0(k_\perp r)}{\tilde{E}^2-k_\perp^2-k_z^2-m^2+i\epsilon}.
    \label{propagadormomentum1}
\end{align}

We perform the integrals over the variables $t$ and $z$ and obtain
\begin{align}
    D(p)&=\int \frac{dEdk_zdk_\perp k_\perp}{2\pi}\delta(p_0-E)\delta(p_z-k_z)\nonumber \\
    &\times \int d\phi dr r \frac{e^{-ip_\perp r \cos{\phi}}J_0(k_\perp r)}{\tilde{E}^2-k_\perp^2-k_z^2-m^2+i\epsilon},
\end{align}
and occupying Eq.~(\ref{J-Aexpansion}) with $l=0$, the propagator becomes
\begin{align}
   D(p)&=\int \frac{dEdk_zdk_\perp k_\perp}{2\pi}\delta(p_0-E)\delta(p_z-k_z)\nonumber \\
    &\times \int dr r \frac{J_0(p_\perp r)J_0(k_\perp r)}{\tilde{E}^2-k_\perp^2-k_z^2-m^2+i\epsilon}.
    \label{propagatorDnospace}
\end{align}
We now use the relation
\begin{equation}
    \int_0^\infty xJ_\alpha(ux)J_\alpha(vx)dx=\frac{1}{u}\delta(u-v),
\end{equation}
in Eq.~(\ref{propagatorDnospace}) and we get
\begin{equation}
     D(p)=\int dE dk_z dk_\perp \frac{\delta(p_0-E)\delta(p_z-k_z)\delta(p_\perp-k_\perp)}{\tilde{E}^2-k_\perp^2-k_z^2-m^2+i\epsilon}, 
\end{equation}
As the last step, in order to get finally the expression of the scalar field propagator in a rotating medium, we integrate over all the remaining variables and get the expression
\begin{equation}
    D(p)=\frac{1}{(p_0+\Omega)^2-p_\perp^2-p_z^2-m^2+i\epsilon},
    \label{finalbosonpropagator}
\end{equation}
where we substitute $\tilde{E}^2=(p_0+\Omega)^2$.

\section{\label{sec3}Effective potential of a hot and rotating system}

With the expression for a scalar field propagator in a rotating medium in Eq.~(\ref{finalbosonpropagator}) at hand, and using the expression for a fermion field propagator reported in Ref.~\cite{Ayala:2021osy}
\begin{align}
    S(p)&=\frac{(p_0+\Omega/2-p_z+ip_\perp)(\gamma_0+\gamma_3)+m(1+\gamma_5)}{(p_0+\Omega/2)^2-\bar{p}^2-m^2+i\epsilon}\mathcal{O}^+ \nonumber \\
    &+ \frac{(p_0-\Omega/2+p_z-ip_\perp)(\gamma_0-\gamma_3)+m(1+\gamma_5)}{(p_o-\Omega/2)^2-\bar{p}^2-m^2+i\epsilon}\mathcal{O}^-,
    \label{finalfermionpropagator}
\end{align} 
we are in the position to study the restoration of the chiral symmetry. For this purpose, we use a simple model called Yukawa model and then we calculate the effective potential at finite temperature and angular velocity, within this model we take into account the 1-loop corrections for both boson and fermion degrees of freedom, ring diagrams correction for boson fields, and finally we build a phase diagram in the temperature-angular velocity plane.

Let's start with a description of the model we use. It is called the Yukawa model which Lagrangian is the following
\begin{align}
    \mathcal{L}=\frac{1}{2}(\partial_\mu \phi)^2+\frac{a^2}{2}\phi^2-\frac{\lambda}{4}\phi^4+i\bar{\psi}\gamma^\mu \partial_\mu \psi-g\bar{\psi}\phi\psi,
    \label{YukawaLagrangian}
\end{align}
where $\phi$ is a real scalar field, $\psi$ is a fermion field with $s=1/2$, the squared mass parameter $a^2$, the self-coupling $\lambda$ and the fermion-boson coupling $g$ are taken to be positive. In order to allow for spontaneous symmetry breaking in this model, we let the $\phi$ field to develop a vacuum expectation value $v$
\begin{equation} 
\phi \rightarrow \phi + v.
\label{shift}
\end{equation}
We understand the vacuum expectation value as the order parameter of the theory. After the shift, we rewrite the Lagrangian as follows
\begin{align}
    \mathcal{L}&=\frac{1}{2}(\partial_\mu \phi)^2+\frac{m_\phi^2}{2}\phi^2-\frac{\lambda}{4}\phi^4+\frac{a^2}{2}v^2-\frac{\lambda}{4}v^4-\lambda v\phi^3\nonumber \\ 
    &-\lambda v^3\phi+i\bar{\psi}\gamma^\mu \partial_\mu \psi-m_f\bar{\psi}\psi-g\bar{\psi}\phi\psi,
    \label{lagrangianwithv}
\end{align}
where $m_\phi^2=3 \lambda v^2-a^2$ and $m_f=g v$ are the dynamical boson and fermion masses, respectively. From Eq.~(\ref{lagrangianwithv}), the tree-level potential can be identifying as
\begin{equation}
    V_{\text{tree}}=-\frac{a^2}{2}v^2+\frac{\lambda}{4}v^4,
\end{equation}
where the tree-level potential develops a minimum, which is called the vacuum expectation value of the $\phi$ field, namely
\begin{equation}
    v_0=\sqrt{\frac{a^2}{\lambda}}.
\end{equation}

Our strategy to study the restoration of the chiral symmetry consists of computing the loop corrections to the tree-level potential within the Matsubara formalism in a thermal field theory. The first corrections corresponds to the one-loop order contribution. This contains two pieces, vacuum and matter. The former is $v$-dependent, therefore when it is added to the three-level potential, the vacuum expectation value changes. In order to avoid such change and to maintain the tree-level vacuum properties, we add counterterms $\delta a^2$ and $\delta \lambda$ requiring that $v_0$ does not change. The one-loop matter contains $T$ and $\Omega$ contributions and corresponds to the mean field approximation for the system’s energy. However, in order to go beyond the mean field approximation we include  corrections that account for the plasma screening effects~\cite{Dolan:1973qd}. Such contributions can be incorporated into the treatment by including the resummation of the ring diagrams~\cite{lebellac:1996,kapusta_gale_2006}.

The effective potential up to ring diagrams order has three contributions, namely
\begin{equation}
    V^{\text{eff}}=V_{\text{b}}^1+V_{\text{f}}^1+V^{\text{ring}}.
    \label{Veffimplicit}
\end{equation}
The general expression for the one-loop boson contribution can be written as
\begin{equation}
    V_{\text{b}}^1=T\sum_{n=-\infty}^{\infty}\int \frac{d^3k}{(2\pi)^3} \ln D(\omega_n,\Omega,\vec{k})^{1/2},
    \label{1-loopbosonic}
\end{equation}
where
\begin{equation}
    D(\omega_n,\Omega,\vec{k})=-\frac{1}{(\omega_n-i\Omega)^2+k_\perp^2+k_z^2+m_\phi^2},
    \label{BosonPropagator}
\end{equation}
is the boson propagator in a rotating system and $\omega_n=2\pi nT$ are the Matsubara frequencies for boson fields. For a fermion field, the general expression for the one-loop correction is
\begin{equation}
   V_{\text{f}}^1=-T\sum_{n=-\infty}^{\infty}\int \frac{d^3k}{(2\pi)^3} \text{Tr}\big[\ln S(\tilde{\omega}_n,\Omega,\vec{k})^{-1}\big],
   \label{1-loopfermionic}
\end{equation}
where
\begin{widetext}
\begin{equation}
    S(\tilde{\omega}_n,\Omega,\vec{k})=-\frac{\big(i\tilde{\omega}_n+\frac{\Omega}{2}-k_z+ik_\perp\big)(\gamma_0+\gamma_3)+m_f(1+\gamma_5)}{\big(\tilde{\omega}_n-i\frac{\Omega}{2}\big)^2+\vec{k}^2+m_f^2}\mathcal{O}^+ 
    - \frac{\big(i\tilde{\omega}_n-\frac{\Omega}{2}+k_z-ik_\perp\big)(\gamma_0-\gamma_3)+m_f(1+\gamma_5)}{\big(\tilde{\omega}_n+i\frac{\Omega}{2}\big)^2+\vec{k}^2+m_f^2}\mathcal{O}^- ,
    \label{fermiomnProp}
\end{equation}
\end{widetext}
is the fermion propagator in a rotating system and $\tilde{\omega}_n=(2n+1)\pi T$ are the Matsubara frequencies for fermion fields. The ring diagrams term is given by
\begin{equation}
    V^{\text{ring}}=\frac{T}{2}\sum_{n=-\infty}^{\infty}\int\frac{d^3k}{(2\pi)^3} \ln{\big(1+\Pi D(\omega_n,\Omega,\vec{k})\big)},
    \label{rings}
\end{equation}
where $\Pi$ is the boson's self-energy. 

The effective potential up to ring diagrams contribution, Eq.~(\ref{Veffimplicit}), as well as the self-energy are computed analytically in the high temperature approximation. This means that we are considering $T>m$, where $m$ can be either the boson or fermion masses, regardless of the relation between $T$ and $\Omega$. The explicit expression for the effective potential is given by
\begin{align}
    V^{\text{eff}}&=-\frac{a^2+\delta a^2}{2}v^2+\frac{\lambda+\delta \lambda}{4}v^4\nonumber\\
    &+\frac{m_\phi^4 }{64 \pi ^2} \bigg[\ln \left(\frac{16 \pi^2T^2}{\mu ^2}\right)-2\gamma_E\bigg]-\frac{\pi ^2 T^4}{90}\nonumber \\
    &+\frac{T^2}{24}\left(m_\phi^2-2 \Omega ^2\right)-\frac{T \left(\Pi+m_\phi^2-\Omega ^2\right)^{3/2}}{12 \pi }\nonumber\\
    &-\frac{\Omega ^2 }{48 \pi ^2}\left(3 m_\phi^2-\Omega ^2\right)\nonumber-\frac{m_f^4}{32 \pi ^2}\bigg[\ln \left(\frac{\pi^2T^2}{\mu ^2}\right)-2\gamma_E\bigg]\nonumber \\
    &-\frac{7 \pi^2 T^4}{360}+\frac{m_f^2T^2}{8} \left(\frac{\Omega ^2}{4 \pi ^2 T^2}+\frac{1}{3}\right)-\frac{T^2 \Omega ^2}{48}-\frac{\Omega ^4}{384},
    \label{Veffexplicit}
\end{align}
where the self-energy $\Pi$, at leading order in the high-$T$ expansion, is given by
\begin{equation}
    \Pi=\lambda \frac{T^2}{4}+g^2 \left(\frac{T^2}{12}+\frac{\Omega ^2}{16\pi^2}\right).
    \label{self-energy}
\end{equation}
The computation of Eqs.~(\ref{1-loopbosonic}), (\ref{1-loopfermionic}), (\ref{rings}) and the self-energy are performed in detail in Appendix~\ref{Apendice}. The result of the calculation can be read in Eq.~(\ref{Veffexplicit}), where we work in the $\overline{MS}$ renormalization scheme, understanding that $\mu$ is the renormalization scale and, as we mentioned before, when the tree-level effective potential is modified by one-loop corrections, the curvature and the position of the minimum are bound to change. The changes are driven from both purely vacuum contributions  and matter effects. The vacuum changes need to be absorbed with a redefinition of the vacuum terms; so, as to make sure that any change in the position of the minimum truly comes from the matter piece. To achieve this goal, we implement \textit{vacuum stability conditions} by introducing two counterterms $\delta a^2$ and $\delta \lambda$, which are fixed from the conditions
\begin{align}
    \left. \frac{1}{2v}\frac{d V^{\text{vac}}}{dv}\right|_{v=v_0}&=0, \nonumber \\
    \left. \frac{d^2 V^{\text{vac}}}{dv^2}\right|_{v=v_0}&=2a^2.
\end{align}
The vacuum effective potential $V^{\text{vac}}$ comes from the limit when $T$ and $\Omega$ go to zero in Eq.~(\ref{Veffexplicit}). After the implementation of this procedure, we get
\begin{align}
    \delta a^2&=-\frac{a^2 }{16 \pi ^2 \lambda } \left[3 \lambda ^2 \ln \left(\frac{2 a^2}{\mu ^2}\right)-2 g^4+6 \lambda ^2\right], \nonumber \\
    \delta \lambda&=\frac{1}{16 \pi ^2}\left[2 g^4 \ln \left(\frac{a^2 g^2}{\lambda  \mu ^2}\right)-9 \lambda ^2 \ln \left(\frac{2 a^2}{\mu ^2}\right)\right].
    \label{counterterms}
\end{align}
\begin{figure}[b]
    \centering
    \includegraphics[scale=0.58]{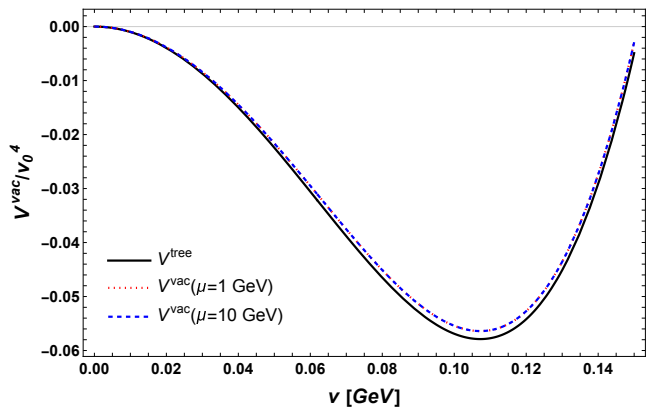}
    \caption{Tree-level potential $V^{\text{tree}}$ and vacuum effective potential $V^{\text{vac}}$ are shown, in order to compare the position and curvature of minimum in both cases. We use $\lambda=0.5$ and $g=0.4$ and for $V^{\text{vac}}$, two different values of the renormalization scale are used, $\mu=1$ and $10$ GeV.}
    \label{fig1}
\end{figure}

Finally, when we substitute Eq.~(\ref{counterterms}) into Eq.~(\ref{Veffexplicit}), the effective potential up to ring diagrams is given by
\begin{align}
    V^{\text{eff}}&=-\frac{a^2}{2}v^2+\frac{\lambda}{4}v^4\nonumber\\
    &+\frac{a^2 v^2}{32 \pi ^2 \lambda } \left[3 \lambda ^2 \ln \left(\frac{2 a^2}{\mu ^2}\right)-2 g^4+6 \lambda ^2\right]\nonumber \\
    &+\frac{v^4}{64 \pi ^2}\left[2 g^4 \ln \left(\frac{a^2 g^2}{\lambda  \mu ^2}\right)-9 \lambda ^2 \ln \left(\frac{2 a^2}{\mu ^2}\right)\right]\nonumber \\
    &+\frac{m_\phi^4 }{64 \pi ^2} \bigg[\ln \left(\frac{16 \pi^2T^2}{\mu ^2}\right)-2\gamma_E\bigg]-\frac{\pi ^2 T^4}{90}\nonumber \\
    &+\frac{T^2}{24}\left(m_\phi^2-2 \Omega ^2\right)-\frac{T \left(\Pi+m_\phi^2-\Omega ^2\right)^{3/2}}{12 \pi }\nonumber\\
    &-\frac{\Omega ^2 }{48 \pi ^2}\left(3 m_\phi^2-\Omega ^2\right)\nonumber-\frac{m_f^4}{32 \pi ^2}\bigg[\ln \left(\frac{\pi^2T^2}{\mu ^2}\right)-2\gamma_E\bigg]\nonumber \\
    &-\frac{7 \pi^2 T^4}{360}+\frac{m_f^2T^2}{8} \left(\frac{\Omega ^2}{4 \pi ^2 T^2}+\frac{1}{3}\right)-\frac{T^2 \Omega ^2}{48}-\frac{\Omega ^4}{384}.    
    \label{Vefffinal}
\end{align}

In order to show that the vacuum contribution to the effective potential does not change the vacuum features, we write the vacuum effective potential taking the limit $T\rightarrow 0$ in Eq.~(\ref{Vefffinal}). Hence, we get
\begin{align}
    V^{\text{vac}}&=-\frac{a^2}{2}v^2+\frac{\lambda}{4}v^4\nonumber \\
    &+\frac{m_\phi^4}{64 \pi ^2} \bigg[\ln \left(\frac{m_\phi^2}{\mu ^2}\right)-\frac{3}{2}\bigg] \nonumber \\
    &-\frac{m_f^4}{32 \pi ^2}\bigg[ \ln \left(\frac{m_f^2}{\mu ^2}\right)-\frac{3}{2}\bigg]\nonumber \\
    &+\frac{a^2 v^2}{32 \pi ^2 \lambda } \left[3 \lambda ^2 \ln \left(\frac{2 a^2}{\mu ^2}\right)-2 g^4+6 \lambda ^2\right]\nonumber \\
    &+\frac{v^4}{64 \pi ^2}\left[2 g^4 \ln \left(\frac{a^2 g^2}{\lambda  \mu ^2}\right)-9 \lambda ^2 \ln \left(\frac{2 a^2}{\mu ^2}\right)\right].
    \label{Veffvac}
\end{align}

In Fig.~\ref{fig1}, we show the tree-level potential $V^{\text{tree}}$ and the vacuum effective potential $V^{\text{vac}}$ computed for $\mu=1$ and $10$ GeV. Notice that, after the vacuum stability conditions are implemented, the vacuum position and curvature remain at their tree-level values and they are independent of the choice of the renormalization scale $\mu$.
\begin{figure}[t]
    \centering
    \includegraphics[scale=0.58]{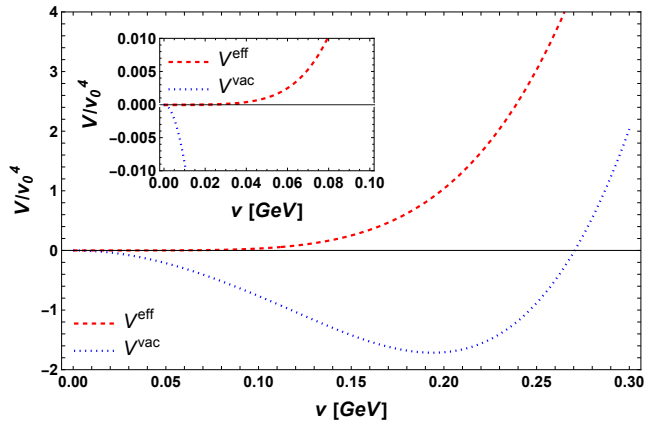}
    \caption{Effective potential (red dashed line) at the phase transition and tree level potential (blue dotted line). Notice the chiral symmetry restoration occurs when $v=0$ and the potential is flat near the minimum. This is tantamount of a second order phase transition. We use the set of parameters $\lambda=1.5$ and $g=0.4$ for both potentials. The critical temperature $T_c=0.498$ GeV and angular velocity $\Omega=0.01$ GeV for the effective potential.}
    \label{fig2}
\end{figure}

We are now in position to explore the restoration of the chiral symmetry, where the temperature is the largest energy scale in the analysis and a rotating system with constant angular velocity $\Omega$ is considered. In the following section, we proceed in this direction.

\section{\label{sec4} Chiral symmetry restoration}

\begin{figure}[t]
    \centering
    \includegraphics[scale=0.58]{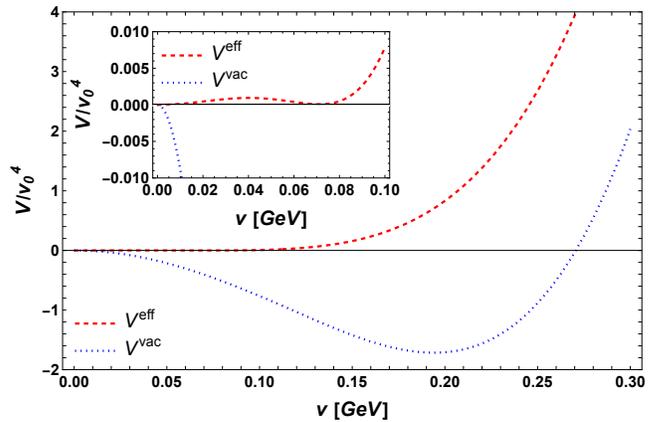}
    \caption{Effective potential (red dashed line) at the phase transition and tree level potential (blue dotted line). Notice the chiral symmetry restoration occurs when the potential develops a hump and two degenerate minima. This is tantamount of a first order phase transition. We use the set of parameters $\lambda=1.5$ and $g=0.4$ for both potentials. The critical temperature $T_c=0.465$ GeV and angular velocity $\Omega=0.1$ GeV for the effective potential.}
    \label{fig3}
\end{figure}

The way in which chiral symmetry restoration is observed occurs when the \textit{vacuum expectation value} is equal to zero. This analysis is carried out through the study of the effective potential Eq.~(\ref{Vefffinal}). We fix the free parameters in the theory, in this case they are $\lambda$ and $g$, then we choose the value of $\Omega$ and vary $T$ until the critical temperature $T_c$ is found, it is determined when the minimum of the effective potential is equal to zero. However, there are two different situations where the chiral symmetry is restored. The first one corresponds to the case when the \textit{vacuum expectation value} is equal to zero, the curvature at this point is equal to zero, so the shape of the potential near the minimum is flat. This is depicted in Fig.~\ref{fig2}. The second case is when there are two degenerate minima, one at zero and other at some finite value of the \textit{vacuum expectation value}, thus the shape of the potential shows a hump between the minima. This is depicted in Fig.~\ref{fig3}.

\begin{figure}[t]
    \centering
    \includegraphics[scale=0.58]{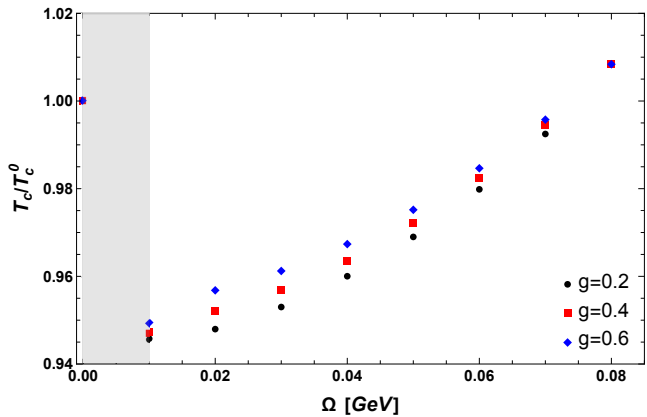}
    \caption{Critical temperature $T_c$ as a function of $\Omega$, using $\lambda=0.4$, $g=0.2, \ 0.4$ and $0.6$. The vertical grey band highlights the region where first order phase transition is found. Outside of the highlighted region the phase transition is of second order.}
    \label{fig4}
\end{figure}

\begin{figure}[b]
    \centering
    \includegraphics[scale=0.58]{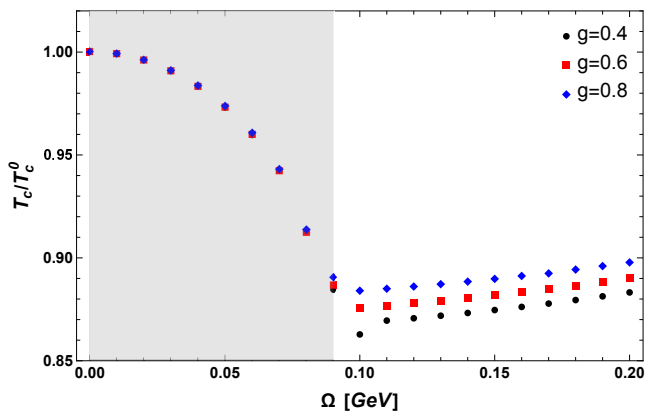}
    \caption{Critical temperature $T_c$ as a function of $\Omega$, using $\lambda=1$, $g=0.4, \ 0.6$ and $0.8$. The vertical grey band highlights the region where first order phase transition is found. Outside of the highlighted region the phase transition is of second order.}
    \label{fig5}
\end{figure}

\begin{figure}[t]
    \centering
    \includegraphics[scale=0.58]{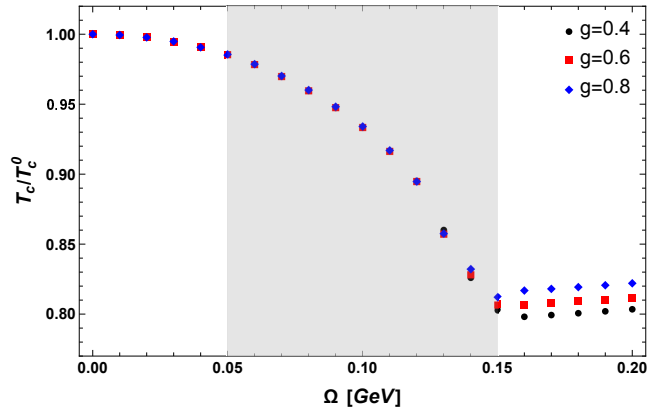}
    \caption{Critical temperature $T_c$ as a function of $\Omega$, using $\lambda=1.5$, $g=0.4, \ 0.6$ and $0.8$. The vertical grey band highlights the region where first order phase transition is found. Outside of the highlighted region the phase transition is of second order.}
    \label{fig6}
\end{figure}

Since, we know how to identify the phase transition and the kind of it as well. We now proceed to analyse the behaviour of $T_c$ as a function of $\Omega$. For this purpose, we choose a pair of sets of values for the free parameters, the coupling constants, $\lambda=0.4, \ 1$ and $1.5$, and $g=0.2, \ 0.4, \ 0.6$ and $0.8$. In order to present the results, we show them in three phase diagrams depicted in Figs.\ref{fig4}-\ref{fig6}. Each of these figures have one value of $\lambda$ and three different values of $g$.

Figure~\ref{fig4} shows the phase diagram obtained for the case $\lambda=0.4$ and $g=0.2, \ 0.4$ and $0.6$, that is, the temperature as a function of $\Omega$, normalised to the critical temperature at $\Omega=0$ GeV. We can notice that the critical temperature $T_c$ decreases abruptly once the angular velocity $\Omega$ is finite. However, in a very fast way $T_c$ reaches its minimum and then it increases monotonically. Furthermore, it is important to notice that from $\Omega=0$ GeV to $\Omega=0.01$ GeV, we observe a first order phase transition, this is highlighted with the shaded vertical band in such region. Outside of the highlighted region, for values of $\Omega>0.01$ GeV, we always find a second order phase transition.

Figure~\ref{fig5} shows the phase diagram obtained when we analyse the temperature as a function of $\Omega$, normalised to the critical temperature at $\Omega=0$ GeV, using $\lambda=1$ and $g=0.4, \ 0.6$ and $0.8$. The behaviour for this case is quite similar to that shown in Fig.~\ref{fig4}. We observe that $T_c$ has a decreasing behaviour from $\Omega=0$ GeV up to $\Omega=0.1$ GeV, where we observe a minimum, from this point onwards the $T_c$ shows an increasing behaviour. We highlight with a shaded vertical band the region where the first order phase transition is found. It coincides with the region where $T_c$ is decreasing. Compared to Fig.~\ref{fig4}, Fig.~\ref{fig5} shows an increase in the region where the first order phase transition is located.

Figure~\ref{fig6} is the last phase diagram reported in this work. Here, we show the behaviour of the normalised temperature as a function of $\Omega$, for $\lambda=1.5$ and $g=0.4, \ 0.6$ and $0.8$. The qualitative behaviour of the $T_c$ is the same as in the other two cases, except that now the region where $T_c$ decreases is larger, it is up to the value $\Omega=0.15$ GeV and the most relevant difference is related to the region where the first order phase transition is located. We notice that the first order phase transition is still contained in the region where the temperature decreases, but for the first time they do not fully coincide. We observe two disconnect second order phase transition regions, the first one at low values of angular velocity, from $\Omega=0$ Gev to $\Omega=0.05$ GeV, and the second region starts at $\Omega>0.15$ GeV, this region at high values of angular velocity shows a gentle monotonic rise in $T_c$. Also, it is important to mention that in Figs.\ref{fig4}-\ref{fig6} the corresponding range in the horizontal axis obeys the high-temperature approximation, in other words, the maximum value of $\Omega$ is not arbitrary, it guarantees $T$ is the highest energy scale.

\section{\label{sec5}Summary}

In this work we have studied the chiral phase transition at finite temperature for a system consisting of fermions and scalars with spontaneous breaking of symmetry and subject to the effects of a rotating system. For the analysis, in Sec.~\ref{sec2}, we have computed the expression corresponding to the scalar field propagator in a rotating medium. We have used the Fock-Schwinger proper-time method and assumed the relativistic system can be described as a rigid cylinder rotating around the $\hat{z}$-axis with constant angular velocity $\Omega$. With the expressions for a scalar and fermion field propagators in a rotating medium at hand. The latter reported in Ref.~\cite{Ayala:2021osy}. In Sec.~\ref{sec3}, we have proceeded to compute all the elements necessary to analyse the restoration of the chiral symmetry. As a first step, we have described the model that we use, which is the Yukawa model, where we have available one scalar field and the fermion and anti-fermion fields. Our next step was to calculate the effective potential beyond the mean field approximation, where we have considered the boson ring diagrams contributions. We have implemented the \textit{vacuum stability conditions} which allow to ensure that the modifications to the shape of the effective potential come only from the matter terms and not from the vacuum contribution. Finally, in Sec.~\ref{sec4} we showed the results. We identified the second and first order first transitions that tell us the chiral symmetry is restored, which is depicted in Figs.~\ref{fig2} and~\ref{fig3}, respectively. Second order phase transition occurs when the minimum of the potential is equal to zero and the potential is flat near the minimum. On the other hand, first order phase transition occurs when  the potential develops a hump between two degenerate minima. 
The last part of this work was devoted to show the change that $T_c$ has when we vary the angular velocity $\Omega$, in other words we built a phase diagram. We show the phase transition lines in the $T-\Omega$ plane. We have fixed values for the couplings $\lambda$ and $g$, which are the free parameters in this theory. For this purpose, we have explored a representative region that gives the different possibilities that exist of values for $\lambda$ and $g$. It means, we have considered the cases when $\lambda<g$, $\lambda \approx g$ and $\lambda >g$. All the combinations between the couplings considered in this work, generating the phase diagrams depicted in Figs.~\ref{fig4}-\ref{fig6}. All the cases have shown that $T_c$ begins with a decreasing behaviour, as a function of $\Omega$, until it reaches a minimum and from there, when increasing $\Omega$, a monotonically increasing behaviour of $T_c$ is obtained. However, we noticed that the region where $T_c$ decreases is greater as $\lambda$ increases its value. From Figs.~\ref{fig4} and~\ref{fig5}, we observed that when the phase transition is first order for zero angular velocity, the latter makes the transition turn into second order, which corresponds to the case when $\lambda<g$ and $\lambda \approx g$. However, for the cases where $\lambda>g$, depicted in Fig.~\ref{fig6}, we noticed that the phase transition is second order for zero angular momentum, eventually changing to a first order phase transition when the angular velocity increases, but if we further increase the angular velocity, we again got a second order phase transition. 

It is quite relevant how the effects coming from a system that is rotating generate changes in the critical temperature and in the type of transition that occurs. Therefore, this work serves as a basis for future work where the effects of a rotating system applied to systems such as relativistic heavy ion collisions are studied, in order to describe the QCD phase diagram through the study of chiral symmetry restoration. This is work for the future and will be reported elsewhere.

\begin{acknowledgments}
I. I. G. acknowledges support from Consejo Nacional de Ciencia y Tecnología Grants No. A1-S-7655 and No. A1-S-16215 R. Z. acknowledges support from ANID/CONICYT FONDECYT Regular (Chile) under Grant No. 1200483.
\end{acknowledgments}

\appendix

\section{\label{Apendice}Effective potential in the high temperature approximation}

We compute explicitly all the corrections up to ring diagrams of the effective potential in the high temperature approximation when the system is rotating. Firstly, we will get the one-loop boson contribution in the high temperature approximation by rewriting the expression in Eq.~(\ref{1-loopbosonic}) as follows
\begin{equation}
    V_{\text{b}}^1=-\frac{T}{2} \sum_{n=-\infty}^{\infty}\int \frac{d^3k}{(2\pi)^3} \ln D(\omega_n,\Omega,\vec{k})^{-1}.
    \label{1-loopbosonicP}
\end{equation}
Substituting the boson propagator in Eq.~(\ref{BosonPropagator}), and implementing the derivative and integral respect to $m_\phi^2$, we get
\begin{equation}
    V_{\text{b}}^1= \frac{T}{2} \int \frac{d^3k}{(2\pi)^3} \int dm_\phi^2 \sum_{n=-\infty}^{\infty} \frac{1}{(\omega_n-i\Omega)^2 + E^2},
    \label{BosPot1}
\end{equation}
where $E=\sqrt{p^2 + m_\phi^2}$, and $\omega_n=2\pi nT$. Now, we perform the sum over the Matsubara frequencies and the one-loop boson contribution becomes
\begin{align}
    V_{\text{b}}^1&=\frac{1}{4} \int \frac{d^3k}{(2\pi)^3} \int dm_\phi^2\frac{1}{E}\bigg[1+ \frac{1}{e^{\beta(E - \Omega)}-1}\nonumber \\
    &+\frac{1}{e^{\beta(E + \Omega)}-1}\bigg],
    \label{V1baftersum}
\end{align}
where $\beta=1/T$. In Eq.~(\ref{V1baftersum}), we observe that the one-loop boson potential can be split into two contributions, the vacuum and the matter one. Hence, $V_{\text{b}}^1$ is rewritten as 
\begin{equation}
    V_{\text{b}}^1= V_{\text{b,vac}}^1 + V_{\text{b,mat}}^1,
\end{equation}
where
\begin{equation}
    V_{\text{b,vac}}^1= \frac{1
}{4} \int \frac{d^3k}{(2\pi)^3} \int dm_\phi^2 \frac{1}{\sqrt{k^2 + m_\phi^2}},
\end{equation}
and
\begin{align}
    V_{\text{b,mat}}^1&= \frac{1}{4} \int \frac{d^3k}{(2\pi)^3} \int dm_\phi^2 \frac{1}{\sqrt{k^2 + m_\phi^2}} \nonumber\\ 
    &\times\bigg[\frac{1}{e^{\beta(\sqrt{k^2 + m_\phi^2} - \Omega)}-1}+ \frac{1}{e^{\beta(\sqrt{k^2 + m_\phi^2} + \Omega)}-1}\bigg].
\end{align}

Integrating each contribution respect to $m_\phi^2$ we obtain the expressions for vacuum and matter contributions
\begin{equation}
    V_{\text{b,vac}}^1= \frac{1}{2} \int \frac{d^3k}{(2\pi)^3} \sqrt{k^2+ m_\phi^2},
\end{equation}
 and 
 \begin{align}
     V_{\text{b,mat}}^1&= \frac{T}{2} \int \frac{d^3k}{(2\pi)^3}  \Big\{\ln \big [1-e^{-\sqrt{k^2+m_\phi^2} +z} \big ]\nonumber\\ 
     &+ \ln \big [1-e^{-\sqrt{k^2+m_\phi^2} -z}\big ] \Big \},
     \label{Bmatter1}
 \end{align}
respectively. The vacuum contribution needs to be regularized and renormalized after the integration over the three dimension momentum. For this purpose, we use the $\overline{MS}$ scheme and then we obtain
\begin{equation}
    V_{\text{b,vac}}^1=- \frac{m_\phi^4}{64\pi^2} \left[ \ln \left( \frac{\mu^2}{m_\phi^2} \right) +\frac{3}{2} \right].
    \label{BosonVacuumContribution}
\end{equation}
Regarding to the matter contribution, first we set the following change of variables
\begin{equation}
x=\frac{k}{T}, \hspace{1cm} y=\frac{m_\phi}{T}, \hspace{1cm} z=\frac{\Omega}{m_\phi}.
\label{changevar}
\end{equation}
So, the expression in Eq.~(\ref{Bmatter1}) becomes
\begin{align}
     V_{\text{b,mat}}^1&= \frac{T^4}{2\pi^2} \int_0^{\infty} x^2dx  \Big \{\ln \big[1-e^{-\sqrt{x^2+y^2} +zy}\big]\nonumber\\ 
     &+ \ln \big[1-e^{-\sqrt{x^2+y^2} -zy}\big] \Big\}.
     \label{Bmatter2}
 \end{align}
It was shown in Ref.~\cite{Haber:1981tr} that the expression in Eq.(\ref{Bmatter2}) is a particular way of 
\begin{align}
    V_{\text{b,mat}}&=\frac{2T^{n+1}}{(4\pi)^{n/2}\Gamma(n/2)}\int_0^{\infty} x^{n-1}dx \nonumber\\
    &\times \Big\{\ln \big [1-e^{-\sqrt{x^2+y^2} +zy}\big] \nonumber\\
    & + \ln \big[1-e^{-\sqrt{x^2+y^2} -zy}\big] \Big \},
    \label{Bmatter3}
\end{align}
which can be written as
\begin{align}
   V_{\text{b,mat}}=&-2\Gamma\bigg[\frac{n+3}{2}\bigg]\bigg(\frac{T}{\sqrt{\pi}}\bigg)^{n+1}\nonumber \\
   &\times \big \{h_{n+2}(y,z) + h_{n+2}(y,-z)\big \}, 
  \label{Bmatter4}
\end{align}
where 
\begin{align}
    h_n(y,z)=\frac{1}{\Gamma(n)} 
\int_0^{\infty} \frac{x^{n-1}}{\sqrt{x^2+y^2}} \left[ \frac{1}{e^{\sqrt{x^2+y^2}-zy}-1} \right]. \nonumber \\
\label{Bmatter5}
\end{align}
The high temperature approximation of Eq.~(\ref{Bmatter2}) becomes into the calculation of $y\ll 1$ at fixed $z$ on the integrals depicted in Eq.~(\ref{Bmatter3}), assuming $z$ is real and $|z|<1$. Such considerations allows to break up the integral $h_n$ in two pieces, for even- and odd-$z$ terms, where the even case is of our interest, since
\begin{equation}
    h_n^e(y,z)=\frac{1}{2}[h_{n}(y,z) + h_{n}(y,-z)].
\end{equation}
Focusing in the case of $n=2 l +1$, that is, when $n$ is odd, the high temperature expansion is found as
\begin{align}
  &h_{2l+1}^e(y,z) = \frac{\pi y^{2l-1}}{2\Gamma (2l+1)}(-1)^l (1-z^2)^{(l-1/2)} \nonumber\\
  &+\frac{(-1)^l}{2[\Gamma (l+1)]^2}\bigg( \frac{y}{2} \bigg)^2 \bigg\{ \ln\bigg( \frac{y}{4\pi}\bigg) + \frac{1}{2}[\gamma-\psi(l+1)] \nonumber\\
 & + l z^2\hspace{0.05cm} _3F_2(1,1,1-l;\frac{3}{2};2;z^2) \bigg\} + \frac{1}{2\Gamma(l+1)} \sum_{k=0}^{l-1}(-1)^k \nonumber\\
& \times \bigg(\frac{y}{2}\bigg)^{2k}\frac{\Gamma(l-k)\zeta(2l-2k)}{\Gamma(k+1)} \hspace{0.05cm} _2F_1(-k,l-k;\frac{1}{2};z^2) \nonumber\\
& + \frac{(-1)^l}{2\Gamma(l+1)}\bigg(\frac{y}{2}\bigg)^{2l}\sum_{k=1}^\infty(-1)^k\bigg( \frac{y}{4\pi}\bigg)^{2k} \nonumber \\
&\times \frac{\Gamma(2k+1)\zeta(2k+1)}{\Gamma(k+1)\Gamma(k+l+1)}\hspace{0.05cm} _2F_1(-k,-l-k;\frac{1}{2};z^2).
\end{align}
We take the case $l=2$ and consider the leading order in the high temperature, $T\gg m_\phi$, and the matter contribution becomes into 
\begin{align}
    V_{\text{b,mat}}^1&= -\frac{\pi^2 T^4}{90}+ \frac{T^2 }{24}(m_\phi^2-2\Omega^2)\nonumber \\
    &-\frac{T}{12\pi}(m_\phi^2-\Omega^2)^{3/2}  -\frac{\Omega^2}{48\pi^2}(3m_\phi^2-\Omega^2)\nonumber \\
    &-\frac{m_\phi^4}{64\pi^2}\bigg[ \ln\bigg( \frac{m_\phi^2}{16\pi^2T^2} \bigg)+2 \gamma_E -\frac{3}{2} \bigg].
    \label{V1bmatter}
\end{align}
Now, we add the Eq.~(\ref{BosonVacuumContribution}) and Eq.~(\ref{V1bmatter}) and the total one-loop boson contribution in the high temperature approximation is
\begin{align}
    V_b^1&=-\frac{\pi ^2 T^4}{90}
    +\frac{T^2}{24}\left(m_\phi^2-2 \Omega ^2\right) \nonumber \\
    &-\frac{T \left(m_\phi^2-\Omega ^2\right)^{3/2}}{12 \pi }
    -\frac{\Omega ^2 }{48 \pi ^2}\left(3 m_\phi^2-\Omega ^2\right)\nonumber\\
    &+ \frac{m_\phi^4 }{64 \pi ^2} \bigg[\ln \left(\frac{16 \pi^2T^2}{\mu ^2}\right)-2\gamma_E\bigg].
    \label{finalV1bcontribution}
\end{align}

Secondly, the one-loop fermion contribution is calculated by substituting the propagator depicted in Eq.~(\ref{fermiomnProp}) into Eq.~(\ref{1-loopfermionic}), and implementing the derivative and integral respect to $m_f$ contribution one gets
\begin{align}
    V_f^1&=-T\sum_{n=-\infty}^\infty  \int dm_f \int \frac{d^3k}{(2\pi)^3} \nonumber\\
    &\times\bigg[ \frac{1}{(\tilde{\omega}_n-i\Omega/2)^2+E^2}+\frac{1}{(\tilde{\omega}_n+i\Omega/2)^2+E^2}\bigg].
    \label{V_f}
\end{align}
In Eq.~(\ref{V_f}), we compute the sum over the Matsubara frequencies and the integral over $m_f$, and we find that the one-loop fermion contribution can also be split into two contributions, the vacuum contribution
\begin{align}
    V_{f,vac}^1=- \int \frac{d^3k}{(2\pi)^3}\sqrt{k^2+m_f^2}, 
\end{align}
and matter contribution
\begin{align}
    V_{f,mat}^1&=- T\int \frac{d^3k}{(2\pi)^3}\nonumber\\
    &\times\bigg\{\ln\bigg[1+e^{-\beta\big(\sqrt{k^2+m_f^2}-\Omega/2\big)} \bigg]\nonumber\\
    &+\ln\bigg[1+e^{-\beta\big(\sqrt{k^2+m_f^2}+\Omega/2\big)}\bigg ]\bigg\}.
    \label{FermionMat1}
\end{align}

The vacuum contribution is straightforward computed and after regularization and renormalization, by using the $\overline{MS}$ scheme, we obtain
\begin{equation}
    V_{f,vac}^1=\frac{m_f^4}{32\pi^2}\bigg[ \ln\bigg( \frac{\mu^2}{m_f^2} \bigg) +\frac{3}{2} \bigg].
    \label{FermionVacuumContribution}
\end{equation}
Meanwhile, the Eq.~(\ref{FermionMat1}) can be written using the change of variables in Eq.~(\ref{changevar}), but with $\bar{z}=\Omega/2T$ instead of $z=\Omega/m_\phi$ and $m_\phi \rightarrow m_f$, thus
\begin{align}
    V_{f,mat}^1&=- \frac{T^4}{2\pi^2} \int x^2dx \nonumber\\
    &\bigg \{\ln\bigg[1 + e^{-\big(\sqrt{x^2+y^2}-\bar{z}\big)} \bigg]\nonumber \\
    &+\ln\bigg[1 + e^{-\big(\sqrt{x^2+y^2}+\bar{z}\big)} \bigg] \bigg\}.
\end{align}

In order to obtain the high temperature expansion, we can rewrite the latest expression as follows
\begin{equation}
    V^1_{f,mat}=-2T^4I_P^e,
\end{equation} 
where 
\begin{equation}
    I_P^e(y,\bar{z})=\frac{1}{2}\bigg[ I_P(y,\bar{z}) + I_P(y,-\bar{z})\bigg],
\end{equation}
corresponds to the even part of 
\begin{equation}
    I_P(y,\bar{z})=1 \int \frac{x^2 dx}{2\pi^2}\ln\bigg(1 + e^{-\big(\sqrt{x^2+y^2}-\bar{z}\big)} \bigg),
    \label{FermionMat2}
\end{equation}
 which matches up with the fermion case studied in Ref.~\cite{Khvorostukhin:2015cha}. In the high temperature expansion, the Eq.~(\ref{FermionMat2}) becomes 
\begin{align}
    I_P(y,\bar{z}&)=-\frac{1}{\pi^2}\bigg\{Li_4(-e^{\bar{z}}) -\frac{y^2}{4}Li_2(-e^{\bar{z}}) \nonumber\\
    & -\frac{y^2}{2}\ln\frac{y}{2}\sum_{n=0}^\infty\frac{1}{n!(n+2)!}\bigg(\frac{y}{2}\bigg)^{2n+2}Li_{-2n}(-e^{\bar{z}})\nonumber\\
    &+\frac{y^2}{2}\sum_{n=0}^\infty\frac{1}{n!(n+2)!}\bigg(\frac{y}{2} \bigg)^{2n+2}\nonumber \\
    &\times \bigg[ \frac{\psi(n+1)+\psi(n+3)}{2}Li_{-2n}(-e^{\bar{z}}) \nonumber \\
    &+\frac{\partial}{\partial s}Li_s(-e^{\bar{z}})\bigg|_{s=-2n} \bigg]
    \bigg\}.
\end{align}
Using the expression above, we find the matter contribution in terms of the polylogarithm functions is
\begin{align}
    &V_{f,mat}^1=\frac{T^4}{\pi^2}[ Li_4(-e^{\Omega/2T}) +Li_4(-e^{-\Omega/2T})]\nonumber\\
    &-\frac{m_f^2T^2}{4\pi^2}[ Li_2(-e^{\Omega/2T}) +Li_2(-e^{-\Omega/2T})]\nonumber\\
    &+\frac{m_f^4}{8\pi^2}\bigg\{\ln\bigg( \frac{m_f}{2T} \bigg)-\frac{\psi(1)+\psi(3)}{2} -\frac{\partial}{\partial s}Li_s(-e^{\Omega/2T})\bigg|_{s=0}\nonumber \\
    &-\frac{\partial}{\partial s}Li_s(-e^{-\Omega/2T})\bigg|_{s=0}\bigg\}-\frac{m_f^2T^2}{2\pi^2}\ln\bigg( \frac{m_f}{2T} \bigg)\nonumber \\
    &\times \sum_{n=1}^\infty\frac{1}{n!(n+2)!}\bigg( \frac{m_f}{2T} \bigg)^{2n+2}\Big [Li_{-2n}(-e^{\Omega/2T})\nonumber\\
   &+Li_{-2n}(-e^{-\Omega/2T})\Big ]+\frac{m_f^2T^2}{2\pi^2}\sum_{n=1}^\infty\frac{1}{n!(n+2)!}\bigg( \frac{m_f}{2T} \bigg)^{2n+2}\nonumber\\
   &\times \bigg[ \frac{\psi(n+1)+\psi(n+3)}{2}\bigg \{ Li_{-2n}(-e^{-\Omega/2T})\nonumber \\
   &+Li_{-2n}(-e^{\Omega/2T}) \bigg \}+\frac{\partial}{\partial s}Li_s(-e^{\Omega/2T})\bigg|_{s=-2n}\nonumber \\
   &+\frac{\partial}{\partial s}Li_s(-e^{-\Omega/2T})\bigg|_{s=-2n} \bigg].
\end{align}
After some simplifications and taking $T\gg m_f$ one gets
\begin{align}
    V_{f,mat}^1&=-\frac{7\pi^2T^4}{360}- \frac{\Omega^2T^2}{48}\nonumber \\
    &-\frac{\Omega^4}{384\pi^2}+\frac{m_f^2}{8}\bigg(\frac{T^2}{3} + \frac{\Omega^2}{4\pi^2}\bigg)\nonumber \\
    &+\frac{m_f^4}{32\pi^2}\bigg[ \ln\bigg( \frac{m_f^2}{\pi^2T^2} \bigg) +2\gamma_E-\frac{3}{4}\bigg].
     \label{FermionMatterContribution}
\end{align}
We add Eq.~(\ref{FermionVacuumContribution}) and Eq.~(\ref{FermionMatterContribution}), and we finally obtain the one-loop fermion contribution
\begin{align}
   V_f^1&= -\frac{7\pi^2T^4}{360} - \frac{\Omega^2T^2}{48}\nonumber\\
   &-\frac{\Omega^4}{384\pi^2}+\frac{m_f^2}{8}\bigg(\frac{T^2}{3}+ \frac{\Omega^2}{4\pi^2}\bigg)\nonumber \\
   &+\frac{m_f^4}{32\pi^2}\bigg[ \ln \bigg(\frac{\mu^2}{\pi^2T^2}\bigg) +2\gamma_E\bigg].
   \label{finalV1fcontribution}
\end{align}

The last term to consider into the effective potential is the ring diagrams contribution, since we are going beyond the mean field approximation. This contribution is written in Eq.~(\ref{rings}) and it can be rewritten as follows
\begin{align}
    V^{\text{ring}}&=\frac{T}{2}\int\frac{d^3k}{(2\pi)^3} \ln{\big(1+\Pi D(\omega_n,\Omega,\vec{k})\big)} \nonumber \\
    &=\frac{T}{4\pi^2}\int dk \ k^2 \big\{ \ln \big(\Omega^2+\vec{k}^2+m_\phi^2+\Pi \big) \nonumber \\
    &-\ln \big(\Omega^2+\vec{k}^2+m_\phi^2\big) \big\},
    \label{Vringmodezero}
\end{align}
where we have considered only the zero Matsubara mode, since it is the dominant contribution in the high temperature approximation and $\Pi$ is the boson's self-energy. We integrate Eq.~(\ref{Vringmodezero}) over the momentum and we obtain
\begin{equation}
   V^{\text{ring}}=-\frac{T}{12\pi}(m_\phi^2-\Omega^2+\Pi)^{3/2} +\frac{T}{12\pi}(m_\phi^2-\Omega^2)^{3/2}. 
   \label{ringsexplicit}
\end{equation}
Then, the effective potential is the result to joint the Eqs.~(\ref{finalV1bcontribution}), (\ref{finalV1fcontribution}) and (\ref{ringsexplicit}), giving as a result the Eq.~(\ref{Vefffinal}).

One piece that we need to know is the expression for the boson's self-energy. It made out of two kind of terms: one corresponds to the boson loop and the other one to the fermion loop. Therefore, the boson's self-energy can be written as follows
\begin{equation}
    \Pi=\Pi_b+\Pi_f,
\end{equation}
where
\begin{equation}
    \Pi_b=\frac{\lambda}{4} \ 12 \ T\sum_{n=-\infty}^\infty \int \frac{d^3k}{(2\pi)^3} D(\omega_n,\Omega,\vec{k}),
    \label{selfenergyboson}
\end{equation}
with $\lambda/4$ as the vertex term and the factor $12$ corresponds to the combinatorial factors obtained from the interaction Lagrangian in Eq.~(\ref{lagrangianwithv}),
and
\begin{equation}
    \Pi_f=-g^2\text{Tr}\big[S(\tilde{\omega}_n,\Omega,\vec{k})S(\tilde{\omega}_n-\omega_m,\Omega, \vec{k}-\vec{p})\big], 
    \label{selfenergyfermion}
\end{equation}
with $g^2$ the two vertices and the trace refers both to the Lorentz and momentum spaces. In order to compute Eq.~(\ref{selfenergyboson}) and Eq.~(\ref{selfenergyfermion}), we notice the relation between the boson contribution to the effective potential and the boson loop in the boson's self-energy
\begin{equation}
    \Pi_b=\frac{\lambda}{4} \ 12 \bigg( 2\frac{d V^1_b}{d m_\phi^2}\bigg),
    \label{selfenergy-potentialboson}
\end{equation}
and the  corresponding relation between the fermion contribution to the effective potential and the fermion loop in the boson's self-energy
\begin{equation}
    \Pi_f=g^2 \bigg( 2\frac{d V^1_f}{d m_f^2}\bigg).
    \label{selfenergy-potentialfermion}
\end{equation}
It is important to mention that Eq.~(\ref{selfenergy-potentialfermion}) is valid only when the external momentum is equal to zero. We know that in general the fermion loop depends on the external momentum, however in the high temperature approximation, we can neglect the external momentum and keep only the leading terms which are $T$ and $\Omega$. Hence, for this work Eq.~(\ref{selfenergy-potentialfermion}) is valid. Finally, we can subtitute Eq.~(\ref{finalV1bcontribution}) into Eq.~(\ref{selfenergy-potentialboson}) and Eq.~(\ref{finalV1fcontribution}) into Eq.~(\ref{selfenergy-potentialfermion}), and adding up the corresponding results, we get
\begin{equation}
    \Pi=\lambda\frac{T^2}{4}+g^2\bigg( \frac{T^2}{12}+\frac{\Omega^2}{16\pi^2} \bigg).
\end{equation}

\bibliography{ourbibliography}

\end{document}